\documentclass[11]{article}

\usepackage{amsfonts}
\usepackage{amsmath}
\usepackage{enumerate}
\usepackage{graphicx}

\begin{document}




\null\vskip-24pt

\hfill KL-TH 00/06

\vskip-10pt

\hfill {\tt hep-th/0006165}

\vskip0.3truecm

\begin{center}

\vskip 3truecm

{\Large\bf

AdS Box Graphs, Unitarity and Operator Product Expansions
}\\ 

\vskip 1.5truecm


{\large\bf L. Hoffmann} \footnote{email:{\tt hoffmann@physik.uni-kl.de}}{\large\bf, L. Mesref }\footnote{email:{\tt mesref@physik.uni-kl.de}}{\large\bf, W.  R\" uhl} \footnote{email:{\tt ruehl@physik.uni-kl.de}}

\vskip 1truecm


{\it Department of Physics, Theoretical Physics\\

University of Kaiserslautern, Postfach 3049 \\

67653 Kaiserslautern, Germany}\\

\end{center}
\vskip 1truecm

\centerline{\bf Abstract}

We develop a method of singularity analysis for conformal graphs which, in particular, is applicable to the holographic image of AdS supergravity theory. It can be used to determine the critical exponents for any such graph in a given channel. These exponents determine the towers of conformal blocks that are exchanged in this channel. We analyze the scalar AdS box graph and show that it has the same critical exponents as the corresponding CFT box graph. Thus pairs of external fields couple to the same exchanged conformal blocks in both theories. This is looked upon as a general structural argument supporting the Maldacena hypothesis.   

\newpage

\section{Introduction}

The AdS/CFT correspondence \cite{maldacena}-\cite{petersen} connects ${\cal N}=4$ supersymmetric $SU(N)$ Yang-Mills theory in four dimensions at large $N$ and strong 't Hooft coupling  $\lambda=g_{YM}^{2}N$ with type $IIB$ supergravity on the $AdS_{5} \times  S^5$ background based on a perturbatively defined action. The correspondence works by comparison of series expansions in powers of $\frac{1}{N^2}$. At leading order many predictions have been verified, and at next order, results such as concerning anomalies, nonrenormalization theorems and $\frac{1}{N^2}$-corrections to field dimensions for composite fields and structure constants of the SYM$_4$ field algebra have been obtained \cite{bilal}-\cite{penati}.

In this context the evaluation of AdS graphs, that represent the holographic image of the AdS perturbation expansion in powers of $\frac{\alpha'}{R^2}=\lambda^{-\frac{1}{2}}$, confronts us with serious technical problems whose difficulty goes much beyond the corresponding CFT flat space graphs. Partly with techniques developed first for CFT in flat space, the exchange graph was calculated and studied in a series of works \cite{liu, hoffmann}. The results of all such calculations were finally expressed in terms of generalized hypergeometric functions. However, in some cases the field dimensions had to be specialized to small natural numbers.

Due to these difficulties, we advocate another approach in this work. We present Green functions as multiple "Mellin-Barnes integrals" \footnote{Inverse Mellin transforms and Barnes integrals are equivalent} over a meromorphic function $\Phi$. This function $\Phi$ is defined as the integral over a positive function on a compact domain. Usually one would expand this integral into a series of ratios of gamma functions, so that $\Phi$ obtains poles from the gamma functions and from the divergence of the series. The latter are difficult to work out \footnote{Except for the functions, say, $_2F_1(1)$ and $_3F_2(1)$ almost nothing is known}. Thus we would like to extract the poles of $\Phi$ by another method. The relevant poles of $\Phi$, namely those to the right of the Mellin-Barnes contours, originate from the divergence (infinity) of the integrand at certain faces or intersections of faces of the regular polyhedral integration domain. So guessing them is not difficult. These poles form sequences which are integrally spaced and tend to $+ \infty$. Of course at the end all Mellin-Barnes integration contours are shifted to $+ \infty$, so that we find series expansions again.

In Section 2 we discuss this method and typical results from the point of view of unitarity of Green functions and operator product expansions. Important information on the structure of the field algebra is obtained this way. Since AdS/CFT correspondence also implies (supposedly) a correspondence between both field algebras (all orders of $\frac{1}{N^2}$ included), the AdS conformal field theory as the holographic picture of supergravity and flat space CFT must therefore already show a partial correspondence on the level of the meromorphic functions $\Phi$. We demonstrate that this is in fact true for the box graph.

In Section 3 we study once again the exchange graph as a simple example of the previously developed method. 

In Section 4 we treat the box graph with arbitrary field dimensions \footnote{This arbitrariness is essential} with our method. We do not give all the details of the lengthy analysis.

A few remarks are added in Section 5.

\section{Critical exponents and unitarity}
\setcounter{equation}{0}
We discuss here the connection between unitarity, operator product expansions and the "critical exponents", that we shall introduce now.
Consider a four-point function in flat space CFT$_d$. 
\begin{figure}[htb]
\begin{centering}
\includegraphics[scale=0.4]{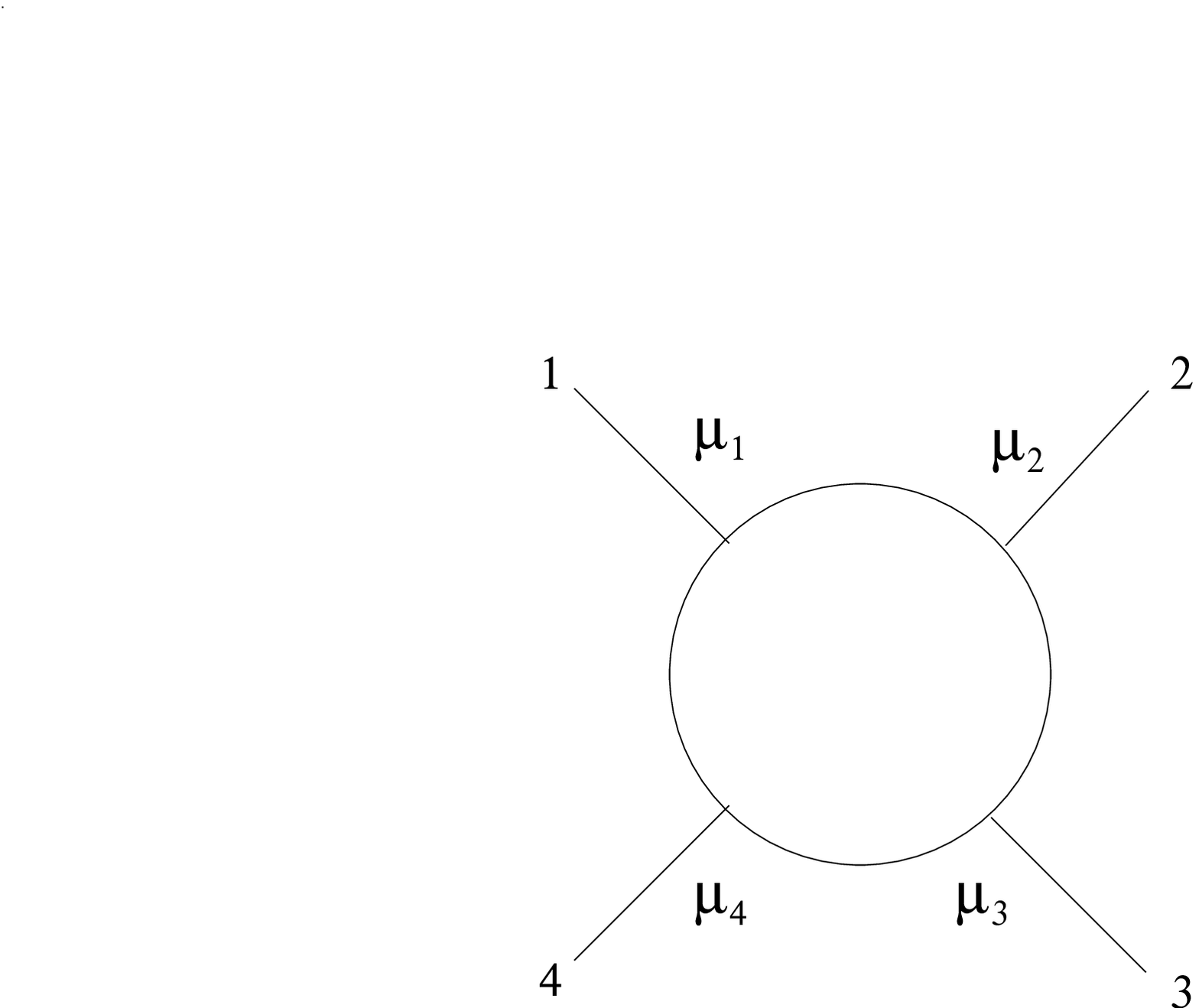}
\caption{An unspecified four-point function of $\text{CFT}_d$}
\end{centering}
\end{figure}
Its Green function $G$ can be split into a covariant multiplier and an invariant function $\tilde{G}$
\begin{equation} \label{2.1}
G(x_1,x_2,x_3,x_4)= (x_{12}^2)^{-\frac{1}{2}(\mu_1+\mu_2-\mu_3+\mu_4)} (x_{13}^2)^{-\frac{1}{2}(\mu_1-\mu_2+\mu_3-\mu_4)}\,(x_{23}^2)^{\frac{1}{2}(\mu_1-\mu_2-\mu_3+\mu_4)} (x_{34}^2)^{-\mu_4}\, \tilde{G}(u,v)
\end{equation}
where
\begin{equation} \label{2.2}
 x_{ij}=x_i-x_j
\end{equation}
and
\begin{equation} \label{2.3}
 u = \frac{x_{14}^2 x_{23}^2}{x_{12}^2 x_{34}^2}, \quad v = \frac{x_{13}^2 x_{24}^2}{x_{12}^2 x_{34}^2}
\end{equation}
are conformal invariant variables. If we intend an operator product expansion in the channel
\begin{equation}
 (1, \, 4) \; \longleftrightarrow \; (2, \,3) \notag
\end{equation}
we must let
\begin{equation} \label{2.4}
 u \rightarrow 0, \quad v \rightarrow 1.
\end{equation}
The function $\tilde{G}(u,v)$ can in turn be decomposed as
\begin{equation} \label{2.5}
 \tilde{G}(u,v)=\sum_{k=1}^{K}\; u^{\gamma_k} \, F_k(u,v),
\end{equation}
where $F_k$ are holomorphic functions in the neighborhood of (\ref{2.4}) and possess the Taylor expansion
\begin{equation} \label{2.6}
 F_k(u,v)=\sum_{m,n=0}^{\infty} \; \frac{u^n \: (1-v)^m}{n! \: m!} \; c_{mn}^{(k)}.
\end{equation}
The $\gamma_k$ are the "critical exponents". Of course, the $\gamma_k$ are, due to possible changes in the covariant multiplier (\ref{2.1}), defined up to a common additive constant. So, what is the physical information encoded in these exponents?

Consider the exchange of the scalar field of dimension $\delta$ in the channel \footnote{We call the exchange channel the "direct" channel}
\begin{equation}
 (1, \, 4) \; \longleftrightarrow \; (2, \,3) \notag
\end{equation}
as described by Fig.2
\begin{figure}[htb]
\begin{centering}
\includegraphics[scale=0.4]{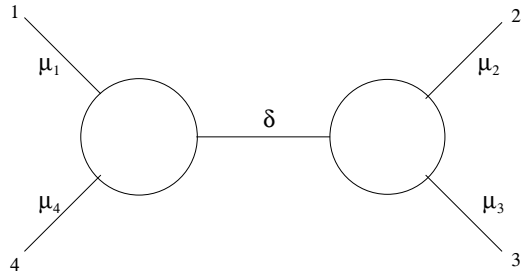}
\caption{Scalar field exchange in the direct channel}
\end{centering}
\end{figure}
where the dimension $\delta$ is assumed to be generic.
Note, that the CFT$_d$ covariant vertex functions
\begin{equation} \label{2.7}
 \int \, dy \, \prod_{i=1}^{n} \, ((y-x_i)^2)^{-\mu_i}
\end{equation}
are necessarily "unique", i.e. they satisfy the condition
\begin{equation} \label{2.8}
  \sum_{i=1}^{n} \: \mu_i \: = \: d.
\end{equation}
A full vertex, such as in Fig.2, can always be resolved in three unique vertices (Fig.3) in an unambiguous fashion.
\begin{figure}[htb]
\begin{centering}
\includegraphics[scale=0.4]{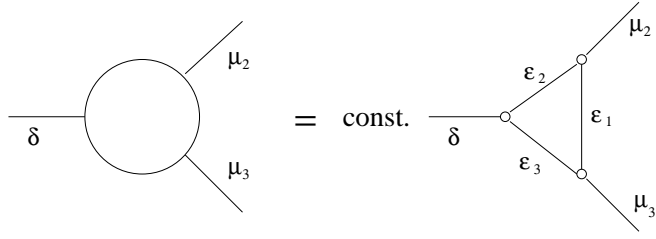}
\caption{Representation of a full three vertex by three unique vertices}
\end{centering}
\end{figure}
This fact can be readily used to compute the Green function corresponding to Fig.2. The result is explicitly known \cite{dobrev} and can be represented as
\begin{equation} \label{2.9}
 \tilde{G}(u, v) = \sum_{k=1}^2 \: u^{\gamma_k} \: F_k(\delta; u, v)
\end{equation}
with
\begin{gather}
 \gamma_1 = \frac{1}{2}(\delta-\mu_1-\mu_4) \label{2.10} \\
 \gamma_2 = \frac{1}{2}(d-\delta-\mu_1-\mu_4) \label{2.11}
\end{gather}
and, after an appropriate renormalization,
\begin{equation} \label{2.12}
 F_2(\delta; u, v) \: = \: F_1(d-\delta; u, v).
\end{equation}
On the other hand, for the holographic image of the AdS exchange graph Fig.4, 
\begin{figure}[htb]
\begin{centering}
\includegraphics[scale=0.4]{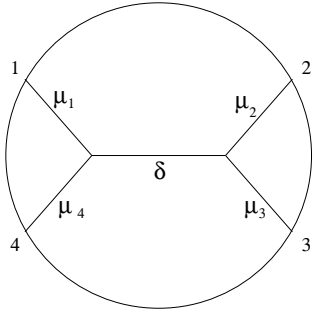}
\caption{The Witten exchange graph}                                    
\end{centering}
\end{figure}
termed "Witten graph", we obtain
\begin{equation} \label{2.13}
 \tilde{G}_{W}(u, v) = \sum_{k=1}^{3} \: u^{\gamma_k} \: F_{W,k}(\delta; u, v)
\end{equation}
with
\begin{align}
 2 \gamma_1 + (\mu_1 + \mu_4)  &= \mu_1 + \mu_4, \label{2.14}  \\
 2 \gamma_2 + (\mu_1 + \mu_4)  &= \mu_2 + \mu_3, \label{2.15}  \\
 2 \gamma_3 + (\mu_1 + \mu_4)  &=  \delta.  \label{2.16}
\end{align}
Of course (\ref{2.10}) and (\ref{2.16}) are identical. On the other hand there are striking differences. 
In CFT jargon the $k=2$ term in (\ref{2.9}) is called "shadow term" of the $k=1$ term. Its appearance is a consequence of conformal harmonic analysis on $\bf{R}_d$ and the equivalence of scalar representations with dimension $\delta$ and $d-\delta$. Only if 
\begin{equation} \label{2.17}
 \delta \leq \frac{d}{2}+1
\end{equation}
a scalar $\underline{\text{field}}$ of dimension $d-\delta$ exists and we have two equivalent formulations of the same CFT$_d$: each external leg of a Green function with dimension $\delta$ can by amputation be transformed into a leg with dimension $d-\delta$ and vice versa.
This shadow term is absent in (\ref{2.13}). Instead, there are two terms $k \in {1, 2}$, which are obviously connected with the exchange of some fields of dimension
\begin{equation}
 \mu_1 + \mu_4 + n  \quad (\mu_2 + \mu_3 + n), \quad n \in \bf{N}_0.\notag
\end{equation}

Now we remember that unitarity of the S-matrix in perturbative quantum field theory is usually formulated by Cutkosky's rule \cite{cutkosky}: cutting a graph (Fig.5)
\begin{figure}[htb]
\begin{centering}
\includegraphics[scale=0.4]{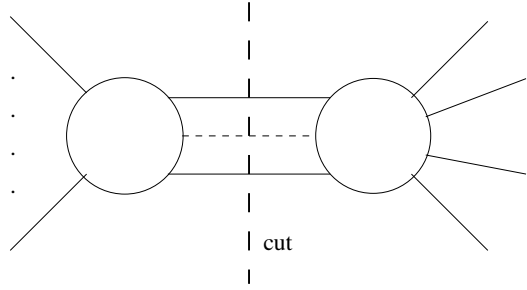}
\caption{ Cutkosky cut through three internal lines of a graph}
\end{centering}
\end{figure}
through internal lines and replacing these by a sum over the corresponding states (with appropriate normalization) gives an absorptive part of the original Green function. In CFT, we can reduce these states by operator product expansion to states created by conformal blocks of fields. In Fig.2 there is one conformal block, namely the conformal field of dimension $\delta$ and all its derivative fields. The same is true for Fig.4 and the part $k=3$, (eqn.(\ref{2.16})). The part $k=1$ (eqn.(\ref{2.14})) involves an infinite number of conformal tensor fields of rank $l$ with dimension 
\begin{equation}
 \mu_1 + \mu_4 + l + 2t \notag
\end{equation} 
and their derivative fields. In fact, there are two parameters $l$ (rank) and $t$ (twist) to label all blocks exchanged. The same is true for $k=2$. The fact that for $k=3$ only one block is exchanged is reflected in the analytic property of $F_{W,3}(u, v)$.
Thus we conclude that each critical exponent corresponds to an infinite tower of conformal blocks, that this tower is determined by a Cutkosky cut acting on internal and external lines and that $2 \gamma_k + \mu_1 + \mu_4$ is in fact the dimension of the lowest dimensional scalar field in the tower, which in turn can be understood as "composite field" of the fields belonging to the lines cut.
Thus the difference between CFT$_{d}$ and AdS$_{d+1}$ theory is in the exchange graphs:
\begin{enumerate}
\item there is no shadow term in AdS$_{d+1}$; 
\item there are terms from cutting external lines in AdS$_{d+1}$. As was argued \cite{hoffmann} in the shadow term in CFT$_d$ and the external line terms in AdS$_{d+1}$ are necessary to guarantee analytic behavior in the crossed channel.  
\end{enumerate}
Indeed, it turns out that such differences between CFT$_d$ and AdS$_{d+1}$ seem to arise $\underline{\text{only}}$ in the exchange graphs \footnote{The usual notation is "one-particle reducible graphs"} in the direct channel.

Next we consider a CFT$_d$ box graph with four unique vertices (Fig.6). The uniqueness conditions imply certain constraints
\begin{figure}[htb]
\begin{centering}
\includegraphics[scale=0.35]{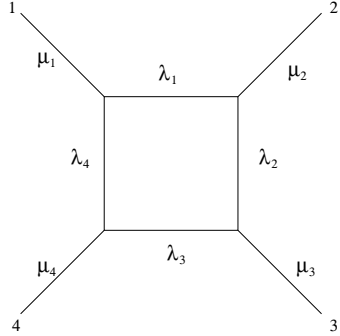}
\caption{The $\text{CFT}_d$ box graph with unique vertices}
\end{centering}
\end{figure}
on the dimensions of the external and internal fields, e.g.
\begin{equation} \label{2.18}
 \mu_1 + \mu_3 = \mu_2 + \mu_4 = 2d - \sum_{i=1}^4 \: \lambda_i.
\end{equation}
This box graph Green function is explicitly known \cite{lang} and
\begin{equation} \label{2.19}
 \tilde{G}(u, v) = \sum_{k=1}^3 \: u^{\gamma_k} \: F_k(u, v)
\end{equation}
with
\begin{align} 
 \gamma_1 &= 0, \label{2.20} \\
 \gamma_2 &= \frac{1}{2}(\lambda_1 + \lambda_3 - \mu_1 -\mu_4), \label{2.21} \\ \gamma_3 &= \frac{1}{2}(\mu_2 + \mu_3 - \mu_1 -\mu_4), \label{2.22}
\end{align}
Now we have Cutkosky cuts through the external pairs of lines as well  $(\gamma_1, \gamma_3)$. We note that the box graph with non-unique vertices (full vertices) has not been calculated yet.
\begin{figure}[htb]
\begin{centering}
\includegraphics[scale=0.3]{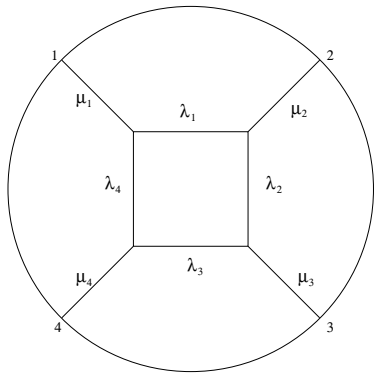}
\caption{The $\text{AdS}_{d+1}$ box graph with generic dimensions}
\end{centering}
\end{figure}
Since the critical exponents determine the towers of conformal blocks that are coupled to the pairs of external and internal fields in the direct channel, they enter the structure of the field algebra. If the Maldacena conjecture in the strong version is correct, the large $\lambda$\, $\frac{1}{N}$-expanded SYM$_d$ with gauge group $SU(N)$ and ${\cal{N}}=4$ supercharges has the same field algebra as the holographic image of the AdS$_{d+1}$ supergravity with coupling constants of order $\frac{1}{N^k}$, $k \in 2 \bf{N}$.
Therefore the results (\ref{2.19}) - (\ref{2.22}) should hold in the case of the Witten graph Fig.7 as well. We shall prove in the sequel, that this is correct indeed.

\section{The singularity analysis of conformally covariant Green functions}
\setcounter{equation}{0}
We aim at a direct determination of the critical exponents $\gamma_k$ (\ref{2.5}) before attempting the explicit evaluation of integral representations. The Taylor coefficients $c_{mn}^{(k)}$ (\ref{2.6}) are then finally represented as integrals which eventually can be evaluated numerically. Since analytic continuation of the integral representations in the parameters (field and space dimension) off the domain of absolute convergence is always tacitly understood, the integrals must necessarily be transformed into absolutely convergent expressions by substraction regularization methods before the numerics can be performed.
The method of analyzing conformal Green functions developed by us consists of several steps:
\begin{enumerate}
\item We derive a multi-parametric Mellin-Barnes  integral representation, where the integrand $\Phi$ depends meromorphically on the Mellin-Barnes parameters and the field and space dimensions. This function $\Phi$ is itself given as an integral of a positive function over a compact polyhedral domain $\Bbb{K}_n$ in $\bf{R}_n$ with possible zeros and infinities on the boundary on $\Bbb{K}_n$. $\Bbb{K}_n$ is the $n$-dimensional generalization of the regular tetrahedron $\Bbb{K}_3$ or the regular triangle $\Bbb{K}_2$. $\Bbb{K}_n$ is bounded by $(n+1)$ faces $\Bbb{K}_{n-1}$, which intersect in edges $\Bbb{K}_{n-2}$ etc.
\item If the integrand is $+ \infty$ on a face or a lower dimensional intersection $\Bbb{K}_r$, then poles may appear in the Mellin-Barnes parameters on the "right" side of the Mellin-Barnes contours. 
\item Two Mellin-Barnes parameters are connected with the kinematical variables $u$ and $1-v$ (\ref{2.3}) by the powers 
\begin{equation} \label{3.1}
 u^{\sigma_1} \, (v-1)^{\sigma_2}.
\end{equation}
The pole positions of $\Phi$ in $\sigma_2$ lie in $\bf{N}_0$ and the shift of the $\sigma_2$ integration contour to $+ \infty$ gives simple power series in $1-v$. The pole positions in $\sigma_1$ lie in different sequences
\begin{equation} \label{3.2}
 \underset{k}{\bigcup} \: \{\gamma_k +\bf{N}_0\}
\end{equation}
which leads us to the series representations (\ref{2.5}), (2.6).
\item Since the zero of an analytically continued integral is difficult to recognize (zeros can $\underline{\text{only}}$ arise after analytic continuation since the original integrand is a positive function) the list of candidates for the exponents $\{\gamma_k\}$ is generally too long. We can reduce this list by different arguments, e.g. a "beta-function argument" and a symmetry argument.
\end{enumerate}
As a nontrivial example of describing our method, we choose the holographic image of the AdS$_{d+1}$ graph in Fig.4. Due to conformal invariance, a Green function can be completely reconstructed if three of its $n \geq 3$ variables are fixed to the values, say
\begin{equation}
 x_1 = 0,\quad x_2 = \infty,\quad x_3 \:\text{arbitrary unit vector} \notag
\end{equation}
We shall exploit this fact by letting $x_3 \rightarrow \infty$, but keeping translational and scale invariance
\begin{equation} \label{3.3}
\underset{x_3 \rightarrow \infty}{\text{lim}} \: (x_3^2)^{\mu_3}\, G(x_1, x_2, x_3, x_4) = (x_{12}^2)^{-\Delta\mu} \, \tilde{G}(u, v)
\end{equation}
with 
 \begin{equation} \label{3.4}
 u= \frac{x_{14}^2}{x_{12}^2}, \quad v=\frac{x_{24}^2}{x_{12}^2}
\end{equation}
and
\begin{equation} \label{3,5}
 \Delta\mu = \frac{1}{2} (\mu_1 + \mu_2 - \mu_3 + \mu_4).
\end{equation}
Denoting the bulk variables of $\text{AdS}_{d+1}$ by $w_1, w_2, w_3, ...$ and boundary variables by $\vec{x}$, $\vec{y}$, $\vec{z}$, ...we have for the bulk-to-boundary propagators $(i \in \{1,2,3,4\})$ \cite{liu}
\begin{equation} \label{3.6}
 K_{\mu{_i}}(w, x_i) = c_{\mu{_i}} \, \left(\frac{w_0}{w_0^2 + (\vec{w}-\vec{x})^2}\right)^{\mu{_i}}
\end{equation}
where
\begin{equation} \label{3.7}
 c_{\mu_i} =\frac{\Gamma(\mu_i)}{\pi^{\frac{d}{2}} \Gamma(\nu_i)}, \quad (\nu_i = \mu_i-\frac{1}{2}d)
\end{equation}	
and
\begin{equation} \label{3.8}
 \underset{x_3 \rightarrow \infty}{\text{lim}} \: (x_3^2)^{\mu_3}\: K_{\mu_3}(w, x_3) = c_{\mu_3} \, w_0^{\mu_3}
\end{equation}		
For the bulk-to-bulk propagator we use the Mellin-Barnes integral representation \cite{liu}
\begin{equation} \label{3.9}
 G_{\lambda}(w,w') = \frac{1}{2 \pi i} \, \int_{-i \infty}^{+i \infty} \, ds \, \Gamma(-s) \, e^{i \pi s} \, \frac{\Gamma(\lambda +2 s)}{\Gamma(\tilde{\nu}+s+1)}\,\frac{1}{2 \pi^{\frac{d}{2}}} \, \left[\frac{w_0 w'_0}{w_0^2 + {w'}_0^2 + (\vec{w} - \vec{w'})^2}\right]^{\lambda + 2s}
\end{equation}	
with $\tilde{\nu} = \lambda - \frac{1}{2}d$.
The graph of interest (Fig.4) is, up to coupling constants, factorials and symmetry factors, represented by the integral
\begin{equation} \label{3.10}
\int \, d\mu(w) d\mu(w') \, G_{\lambda}(w, w') \underset{i \in \{1, 4\}}{\prod} K_{\mu_i}(w, x_i) \, \underset{j \in \{2, 3\}}{\prod} K_{\mu_j}(w', x_j)
\end{equation}
where $d\mu$ is the invariant $\text{AdS}_{d+1}$ measure
\begin{equation} \label{3.11}
 d\mu(w) = \frac{dw_0 \, d\vec{w}}{w_0^{d+1}}
\end{equation}

The integration starts by using a $\Gamma$-function auxiliary integration for each denominator in (\ref{3.6}), (\ref{3.9})
\begin{equation} \label{3.12}
 \frac{1}{\parallel x \parallel^{2\mu}} = \frac{1}{\Gamma(\mu)} \, \int_0^\infty \, dt \, t^{\mu-1} e^{-t \parallel x \parallel^2}
\end{equation}
distributing the parameters $\{t_i\}_{1,2,4}$ to the $K_{\mu_i}$ and $r$ to $G_\lambda$. Then $w_0$ and $w'_0$ can be integrated giving 
\begin{equation} \label{3.13}
 \underset{i \in\{1, 2\}}{\prod} \, \frac{1}{2} \Gamma(\frac{1}{2} \alpha_i) \, \eta_i^{-\frac{1}{2}\alpha_i}
\end{equation}
with 
\begin{equation} \label{3.14}
\eta_1 = r + t_1 + t_4, \quad \eta_2 = r + t_2
\end{equation}
and 
\begin{equation} \label{3.15}
\alpha_1 = \mu_1 + \mu_4 + \lambda + 2s - d, \quad
\alpha_2 = \mu_2 + \mu_3 + \lambda + 2s - d
\end{equation}
The $\vec{w}, \vec{w}'$ integration is Gaussian and gives
\begin{equation} \label{3.16}
 (\frac{\pi^2}{\text{det}A})^{\frac{d}{2}} \text{exp}\{\chi^T A^{-1} \chi - D\}
\end{equation}
with
\begin{align}
 A &= \left(\begin{matrix} \eta_1 & -r \\ -r & \eta_2 \end{matrix}\right) \label{3.17}, \\
 D &= \sum_i \, t_i x_i^2, \label{3.18} \\
 \chi &= \left(\begin{matrix} t_1 x_1 + t_4 x_4 \\ t_2 x_2 \end{matrix}\right) \label{3.19}
\end{align}
The exponent (\ref{3.16}) can be written as a quadratic form
\begin{equation} \label{3.20}
 -\frac{1}{\text{det}A} \, \sum_{i < j} \, \beta_{ij} (x_i - x_j)^2
\end{equation}
where
\begin{equation} \label{3.21}
 \beta_{12} = r t_1 t_2, \: \beta_{14} = \eta_2 t_1 t_4, \: \beta_{24} = r t_2 t_4
\end{equation}
Since we aim at an expression of the type (\ref{2.6}), we can use (\ref{3.4}) to write (\ref{3.20}) as
\begin{equation} \label{3.22}
 -x_{12}^2 \frac{\beta_0}{\text{det}A} \, \{1 + \frac{\beta_{14}}{\beta_0} u +\frac{\beta_{24}}{\beta_0} (v-1)\}
\end{equation}
with
\begin{equation} \label{3.23}
 \beta_0 = \beta_{12} + \beta_{24} = r t_2 (t_1 + t_4)
\end{equation}
Following Symanzik \cite{symanzik}, the second and third term in (\ref{3.22}) are represented by Mellin-Barnes integrals
\begin{equation} \label{3.24}
 e^{-x} = \frac{1}{2 \pi i} \, \int_{-i \infty}^{+i \infty} \, d\sigma \, \Gamma(-\sigma) \, x^\sigma
\end{equation}
Finally we perform one integration by introducing scaled parameters
\begin{equation} \label{3.25}
 T = r + \sum_{i \in \{1, 2, 4\}} t_i, \quad t_i = T \tau_i, \: i \in \{1, 2, 4\}, \quad r = T \rho,
\end{equation}
so that 
\begin{equation} \label{3.26}
 \text{det}A = T^2 \, [\rho (1 - \rho) + \tau_2 ( 1- \rho - \tau_2)].
\end{equation}
The remaining parameter integrals can then be summed up into a meromorphic function.
\begin{align} \label{3.27}
\Phi(\sigma_1, \sigma_2, s) &= \Gamma(-\sigma_1) \Gamma(-\sigma_2) \Gamma(-s) \int_{\Bbb{K}_2} \, d\tau_1 d\tau_2 d\tau_4 d\rho \, \delta(1-\tau_1-\tau_2 -\tau_4 -\rho) \tau_1^{\mu_1-1} \tau_2^{\mu_2-1} \tau_4^{\mu_4-1} \rho^{\lambda+2s-1} \nonumber \\
&\times (1-\tau_2)^{-\frac{1}{2} \alpha_1} (\rho + \tau_2)^{-\frac{1}{2}\alpha_2} [\rho \tau_2(\tau_1+\tau_4)]^{-\Delta\mu -\sigma_2} \,[\tau_1 \tau_4 (\rho +\tau_2)]^{\sigma_1} (\frac{\tau_4}{\tau_1+\tau_4})^{\sigma_2} \notag \\ &\times [\rho (1-\rho) + \tau_2 (1-\rho-\tau_2)]^{-\frac{1}{2}d+\Delta\mu}
\end{align}
and this enters a threefold Mellin-Barnes integral
\begin{align} \label{3.28}
\tilde{G}(u, v) &= \frac{1}{8 \pi^{\frac{3}{2}d}} \, \frac{\Gamma(\mu_3)}{\prod_{i=1}^4 \, \Gamma(\nu_i)} \, (2\pi i)^{-3} \iiint\limits_{-i\infty}^{+i\infty} \, d\sigma_1 d\sigma_2 ds \,\frac{\Gamma(\frac{1}{2}\alpha_1) \Gamma(\frac{1}{2}\alpha_2) \Gamma(\Delta\mu +\sigma_2 +\sigma_2)}{\Gamma(\tilde{\nu}+s+1)}\notag \\ &\times e^{i \pi s} \, u^{\sigma_1} \, (v-1)^{\sigma_2} \, \Phi(\sigma_1, \sigma_2, s)
\end{align}
with $\Delta\mu = \frac{1}{2} (\mu_1 + \mu_2 - \mu_3 + \mu_4)$, see (3.5). Such a representation (\ref{3.27}), (\ref{3.28}) of any four-point function for $\text{CFT}_d$ or $\text{AdS}_{d+1}$ field theory is the starting point for our singularity analysis, leading to the critical exponents.

In this particular case we can simplify the integral representation (\ref{3.27}) by integrating over $\xi$ in 
\begin{equation} \label{3.29}
 \tau_1 = \tau \xi, \: \tau_4 = \tau (1-\xi)
\end{equation}
\begin{align} \label{3.30}
 \Phi(\sigma_1, \sigma_2, s) &= \frac{\Gamma(\mu_1 +\sigma_1) \Gamma(\mu_4 + \sigma_1 +\sigma_2)}{\Gamma(\mu_1 + \mu_4 + 2\sigma_1 + \sigma_2)} \, \Gamma(-\sigma_1) \Gamma(-\sigma_2) \Gamma(-s)\, \int_{\Bbb{K}_{2}} \, d\tau d\tau_2 d\rho \delta(1-\tau-\tau_2-\rho) \nonumber \\
&\times \tau^{\mu_1+\mu_4-\Delta\mu+\sigma_1-1} \, (1-\tau)^{-\frac{1}{2} \alpha_2 +\sigma_1} \, \tau_2^{\mu_2-\Delta\mu-\sigma_1-1} (1-\tau_2)^{-\frac{1}{2} \alpha_1} \nonumber \\
&\times \rho^{\lambda +2s-\Delta\mu-\sigma_1-1} [\rho (1-\rho) +\tau_2 (1-\rho-\tau_2)]^{-\frac{1}{2}d+\Delta\mu}
\end{align}
Here $\sigma_2$ has vanished from the integral into the factor in front. Except for the factor $\Gamma(-\sigma_2)$, there is no pole to the right of the $\sigma_2$ Mellin-Barnes contour. This a general feature since (see(\ref{3.23})) in $\Bbb{K}_n$
\begin{align} \label{3.31}
 0 \:\leq \:\frac{\beta_{24}}{\beta_0} = \frac{\beta_{24}}{\beta_{12}+\beta_{24}} \:\leq \:1
\end{align}
There are obviously poles from the faces $\tau_2=0$ and $\rho=0$ in $\sigma_1$, arising from the Mittag-Leffler expansion 
\begin{equation} \label{3.32}
 t^{\mu-1} \Theta(t) \underset{\text{poles only}}{\cong} \sum_{n=0}^\infty \, \frac{(-1)^n \delta^{(n)}(t)}{n! (\mu + n)}
\end{equation}
with positions $-\mu \in \bf{N}_0$. Including the poles in $\sigma_1$ from the factor $\Gamma(-\sigma_1)$, we have three possibilities: $(n \in \bf{N}_0)$	
\begin{enumerate}
\item 
\begin{equation} \label{3.33}
        \sigma_1 = n
\end{equation}
\item 
\begin{equation} \label{3.34}
        \text{from}\quad \tau_2 = 0: \quad \sigma_1 = \mu_2 - \Delta\mu +n
     \end{equation}
\item
\begin{equation} \label{3.35}
        \text{from} \quad\rho = 0: \quad \sigma_1 = \lambda +2s- \Delta\mu +n
     \end{equation}
\end{enumerate}
In the cases (1.) and (2.) we get the critical exponents
\begin{align}
  \gamma_1 &= 0, \label{3.36} \\
  \gamma_2 &= \frac{1}{2} (\mu_2 + \mu_3 - \mu_1 - \mu_4) \label{3.37}
\end{align}
whereas case (3.) necessitates knowledge of the pole positions in $s$. One possibility is that these poles are produced by $\Gamma(-s)$, then
\begin{align}
  \sigma_1 &= \lambda- \Delta\mu+n, \label{3.38} \\
  \gamma_3 &= \lambda -\Delta\mu \label{3.39}
\end{align}
There is another candidate for poles in $\sigma_1$, namely the intersection of the faces (2) and (3):
\begin{equation} \label{3.40}	
 \tau_2 = \rho = 0, \quad\tau = 1.	
\end{equation}	
We use the parameters	
\begin{equation} \label{3.41}	
 \rho = \omega\psi, \quad \tau_2 = \omega(1-\psi), \quad \tau = 1-\omega
\end{equation}	
The behavior of the integrand at $w \rightarrow 0$ is given by	
\begin{equation} \label{3.42}	
 \int_0 \: d\omega \,\omega^{(\mu_2-\Delta\mu-\sigma_1) + (\lambda +2s-\Delta\mu-\sigma_1)+(-\frac{1}{2}\alpha_2+\sigma_1)+(-\frac{1}{2}d+\Delta\mu)-1}	
\end{equation}	
The exponent is 
\begin{equation} \label{3.43} 
 \frac{1}{2}\lambda +s-\frac{1}{2}(\mu_1+\mu_4)-\sigma_1-1 
\end{equation}  
and gives rise to poles in $\sigma_1$ at 
\begin{enumerate}
\item[4.]	
\begin{equation} \label{3.44}	
        \tau_2 = \rho = 0: \quad \sigma_1 = \frac{1}{2} \lambda +s- \frac{1}{2} (\mu_1+\mu_4) + n 
    \end{equation}
\end{enumerate} 
If the $s$ poles are from $\Gamma(-s)$, we get from (\ref{3.44})
\begin{align}
 \sigma_1 &= \frac{1}{2} (\lambda - \mu_1 - \mu_4)+n , \label{3.45} \\
 \gamma_4 &= \frac{1}{2}(\lambda-\mu_1-\mu_4)
\end{align}

But there exist other $s$-poles. If we consider (\ref{3.44}), set $n = 0$ and insert the delta function following from (\ref{3.32}) into (\ref{3.30}), there remains the $\psi$-integral, see (\ref{3.41})
\begin{multline} \label{3.47}
 \int_0^1 d\psi\, \psi^{\lambda +2s-\Delta\mu-\sigma_1-1} (1-\psi)^{\mu_2-\Delta\mu-\sigma_1-1} \arrowvert_{\sigma_1 = \frac{1}{2}\lambda+s-\frac{1}{2}(\mu_1+\mu_4)} \\
= \frac{1}{\Gamma(\mu_3)}\Gamma(\frac{1}{2}(\mu_2+\mu_3-\lambda)-s) \Gamma(\frac{1}{2}(\lambda+\mu_3-\mu_2)+s)
\end{multline}
which shows, that there exist relevant $s$-poles from the first factor in the numerator. For arbitrary $n$ in (\ref{3.44}) the poles lie at
\begin{equation} \label{3.48}
 s + n = \frac{1}{2}(\mu_2 + \mu_3 - \lambda) + n', \quad n' \in \bf{N}_0
\end{equation}

If we consider (\ref{3.35}) at $n = 0$ and insert it together with the delta function (\ref{3.32}) into (\ref{3.30}), then the integral turns into a beta-function
\begin{multline}\label{3.49}
\int_0^1 d\tau_2 \tau_2^{(\mu_2-\Delta\mu-\sigma_1-1)+(-\frac{1}{2}\alpha_2+\sigma_1)+(-\frac{1}{2}d+\Delta\mu)} \,(1-\tau_2)^{(\mu_1+\mu_4-\Delta\mu+\sigma_1-1)-\frac{1}{2}\alpha_1+(-\frac{1}{2}d+\Delta\mu)} \\
= \frac{1}{\Gamma(0)}\Gamma(\frac{1}{2}(\mu_2-\mu_3-\lambda)-s) \Gamma(\frac{1}{2}(\lambda-\mu_2+\mu_3)+s)
\end{multline}
The denominator is unchanged if we let $n$ in (\ref{3.35}) assume arbitrary values from $\bf{N}_0$. Thus the denominator of the beta-function lets the singularity (\ref{3.35}) vanish, implying that (\ref{3.38}), (\ref{3.39}) do not exist either. Only in exceptional cases do we get control over the zeros when we can perform an integral completely. Often the integral is a beta-function, then we call our way of proof "the beta-function argument". More effort is needed to evaluate integrals in terms of functions $_{p+1}F_p(1)$ in which case the zeros are also controllable.  

A simple but surprisingly powerful argument to eliminate whole sequences of poles comes from the symmetry of the graph (Fig.4). We define this symmetry to consist of those mappings of the graph on itself:
\begin{enumerate}
\item[(a)] which lead to the same graph after an appropriate relabelling of the external coordinates and the field dimensions;
\item[(b)] leave $u$ and $v$ invariant.
\end{enumerate}
In the case of Fig.4, this leads to a group $Z_2 \times Z_2$, generated by the reflections
\begin{align} \label{3.50}
 S_1 &: 1  \longleftrightarrow  2, \: 3  \longleftrightarrow  4 \notag \\
 S_2 &: 1  \longleftrightarrow  4, \: 3  \longleftrightarrow  2
\end{align}
While the Green function $G(x_1, x_2, x_3, x_4)$ is invariant under $Z_2 \times Z_2$ by definition, the invariant function $\tilde{G}$ is not. Let $S_i(\tilde{G})$ denote the function obtained by applying (\ref{3.50}) to the dimensions in $\tilde{G}$, then from (\ref{2.1}) we obtain
\begin{equation} \label{3.51}
 \tilde{G}(u, v) = u^{\delta_i} v^{\epsilon_i} S_i(\tilde{G})(u, v)
\end{equation}
with
\begin{alignat}{2}
 S_1 :\: \delta_1 &= \frac{1}{2}(\mu_2+\mu_3-\mu_1-\mu_4), \: \epsilon_1 &= \frac{1}{2}(\mu_1+\mu_3-\mu_2-\mu_4)  \label{3.52}\\
 S_2 :\: \delta_2 &= 0, \:\epsilon_2 &= \epsilon_1 \label{3.53}
\end{alignat} 
Inserting (\ref{3.51}) into (\ref{2.5}), we see that the labels $\{k\}$ of $\gamma_k$ are submitted to a representation of $Z_2 \times Z_2$: $S_i \rightarrow \sigma_i$, so that:
\begin{align}
 S_i(\gamma_k)+\delta_i &= \gamma_{\sigma_i(k)} \label{3.54} \\
 S_i(G_k) v^{\epsilon_i} &= G_{\sigma_i(k)} \label{3.55}
\end{align}
Holomorphy of $G_k$ at $v=1$ is obviously not touched by (\ref{3.55}).
Applying (\ref{3.54}) to the graph Fig.4, we find
\begin{alignat}{2} \label{3.56}
 \sigma_1(1) &= 2; \quad \sigma_1(2) &= 1, \nonumber \\
 \sigma_2(1) &= 1; \quad \sigma_2(2) &= 2
\end{alignat}
and 
\begin{equation} \label{3.57}
 \sigma_{1,2}(4) = 4
\end{equation}
whereas $\gamma_3$ does not fit into any representation.

\section{The AdS box graph}
\setcounter{equation}{0}
 Now we turn to the box graph Fig.7. In terms of bulk-to-bulk propagators $G_\lambda$ and bulk-to-surface propagators $K_\mu$, the Green function is given by the integral
\begin{equation}	\label{4.1}
 G(x_1,x_2, x_3, x_4) = \int \, \prod_{i=1}^4 \, d\mu(w_i) \, K_{\mu_i}(x_i, w_i) \, G_{\lambda_i}(w_i, w_{i+1}), \quad w_5 = w_1
\end{equation}
Again we consider the limit (\ref{3.3}), (\ref{3.8}). Due to the four bulk-to-bulk propagators, the invariant Green function $\tilde{G}(u, v)$ has the form of a sixfold Mellin-Barnes integral\footnote{All the techniques and notations used are the same as in the preceeding section.}
\begin{align}	\label{4.2}
\tilde{G}(u, v) &= \frac{1}{2^8 \pi^{2d}} \frac{\Gamma(\mu_3)}{\prod_{i=1}^4 \,\Gamma(\nu_i)} \, (2\pi i)^{-6} \, \iint\limits_{-i\infty}^{+i\infty} \, d\sigma_1 d\sigma_2 \,\iiiint\limits_{-i\infty}^{+i\infty}\, \{\prod_{i=1}^4 \, ds_i \, \frac{\Gamma(\frac{1}{2}\alpha_i)}{\Gamma(\tilde{\nu}_i+ s_i +1)}\} \notag \\ &\times \Gamma(\Delta\mu +\sigma_1 +\sigma_2) \,e^{i\pi\sum_is_i}\, u^{\sigma_1}\, (v-1)^{\sigma_2}\, \Phi(\sigma_1, \sigma_2, s_1, s_2, s_3, s_4)
\end{align}
where
\begin{equation} \label{4.3}
\alpha_i = \mu_i + \lambda_i + 2s_i + \lambda_{i-1} + 2s_{i-1} - d, \quad (\lambda_0 = \lambda_4,\: s_0 = s_4)
\end{equation}
and the meromorphic function $\Phi$ is given by
\begin{multline}\label{4.4}
\Phi(\sigma_1, \sigma_2, s_1, s_2, s_3, s_4) = \prod_{i=1}^2 \, \Gamma(-\sigma_i) \, \prod_{j=1}^4 \, \Gamma(-s_j) \, \int_{\Bbb{K}_6} \, (\prod_{i=1(\neq3)}^4 \, d\tau_i \, \tau_i^{\mu_i-1}) (\prod_{j=1}^4 \, d\rho_j \, \rho_j^{\lambda_j+2s_j-1}\epsilon_j^{-\frac{1}{2}\alpha_j}) \nonumber \\
\times \,\delta(1-\sum_{i=1(\neq3)}^4 \tau_i-\sum_{j=1}^4 \rho_j) \,\delta(A)^{-\frac{d}{2}+\Delta\mu}\,f_0^{-\Delta\mu-\sigma_1-\sigma_2} \,f_1^{\sigma_1} \,f_2^{\sigma_2}
\end{multline}
Here
\begin{equation} \label{4.5}
\epsilon_i = \tau_i + \rho_i + \rho_{i-1}, \quad (\tau_3 =0 ,\: \rho_0 = \rho_4)
\end{equation}
and the remaining functions $f_0, f_1, f_2$ and $\delta(A)$ can be represented best with the help of elementary symmetric polynomials
\begin{align} \label{4.6}
S_2(1, 2, 3) &= \rho_1 \rho_2 + \rho_1 \rho_3 + \rho_2 \rho_3 \notag    \\
S_3(1, 2, 3, 4) &= \rho_1 \rho_2 \rho_3 + \rho_1 \rho_3 \rho_4 + \rho_2 \rho_3 \rho_4 + \rho_1 \rho_2 \rho_4
\end{align}
namely 
\begin{align}
f_0 &= \tau_2[\tau_1 \tau_4 S_2(1,2,3) + (\tau_1 + \tau_4) S_3(1,2,3,4)] \label{4.7} \\
f_1 &= \tau_1 \tau_4[S_3(1,2,3,4) + \tau_2 \rho_4(\rho_2+\rho_3)] \label{4.8} \\
f_2 &= \tau_2 \tau_4[S_3(1,2,3,4) + \tau_1 \rho_2 \rho_3] \label{4.9} \\
\delta(A) &= \tau_1 \tau_2 \tau_4 (\rho_2 + \rho_3) + \tau_1 \tau_2 S_2(2,3,4) + \tau_1 \tau_4 S_2(1,2,3) \notag    \\
&+ \tau_2 \tau_4 (\rho_1 + \rho_4)(\rho_2 + \rho_3) + (\tau_1 + \tau_2 + \tau_4) S_3(1,2,3,4) \label{4.10}
\end{align}
The function $\delta(A)$ originates from the determinant in the Gaussian integration. It is obvious that
\begin{equation} \label{4.11}
0 \:\leq \:\frac{f_2}{f_0}\: \leq\: 1 \quad\text{on}\: \Bbb{K}_6
\end{equation}
so that the only relevant poles in $\sigma_2$ arise from $\Gamma(-\sigma_2)$.

The analysis of the pole positions in $\sigma_1$ is rather involved. In the sequel $n_0 \in \bf{N}_0$ holds throughout. There is one face of type $\Bbb{K}_5$ producing a singularity:
\begin{enumerate}
\item[I]    $\tau_2 = 0$: poles appear at
\begin{align}
\sigma_1 &= \frac{1}{2}(\mu_2+\mu_3-\mu_1-\mu_4) + n_0 \label{4.12} \\
\gamma_1 &= \frac{1}{2}(\mu_2+\mu_3-\mu_1-\mu_4)
\end{align}
\end{enumerate}
There are two faces of $\Bbb{K}_4$ type leading to poles.
\begin{enumerate}
\item[II]
     $\rho_1=\rho_2=0$: we introduce the parameters
\begin{equation}	\label{4.14}
\rho_1= \rho \xi, \quad \rho_2 = \rho (1-\xi)
\end{equation}
and let $\rho \rightarrow 0$. This gives pole positions 
\begin{equation}	\label{4.15}
\sigma_1 = \lambda_1 + \lambda_2 + 2(s_1+s_2)-\Delta\mu +n_0
\end{equation}
If the pole positions of $s_1, s_2$ are chosen from $\bf{N}_0$, we get
\begin{equation}	\label{4.16}
\gamma_2 = \lambda_1 + \lambda_2 - \Delta\mu
\end{equation}
\end{enumerate}
The other case is
\begin{enumerate}
\item[III]  
$\rho_1 = \rho_3 = 0$: this case is treated analogously to case (II). We find poles at
\begin{equation}	 \label{4.17}
\sigma_1 = \lambda_1 + \lambda_3 + 2(s_1+ s_3) - \Delta\mu +n_0
\end{equation}
If the pole positions of $s_1, s_3$ are from $\bf{N}_0$, we find
\begin{equation}	  \label{4.18}
\gamma_3 = \lambda_1 + \lambda_3 - \Delta\mu
\end{equation}
\end{enumerate}
Now we come to the intersections $\Bbb{K}_5 \cap \Bbb{K}_4$	and    $\Bbb{K}_4 \cap \Bbb{K}'_4$ of $\Bbb{K}_3$ type.
\begin{enumerate}
\item[IV]
     $\rho_1 = \rho_2 = \rho_3 = 0$: we choose as parameters
\begin{equation}	  \label{4.19}
\rho_i = \rho\xi_i,\: i \in \{1,2,3\}, \sum_i \,\quad \xi_i = 1
\end{equation}
and let $\rho \rightarrow 0$. Pole positions are
\begin{equation}	  \label{4.20}
\sigma_1 = \lambda_1 + \frac{1}{2}(\lambda_2+\lambda_3)+2s_1+s_2+s_3-\frac{1}{2}(\mu_1+\mu_2+\mu_4) +n_0
\end{equation}
If the $\{s_i\}_{i=1}^3$ have poles in $\bf{N}_0$, we get
\begin{equation}	  \label{4.21}
\gamma_4 = \lambda_1 + \frac{1}{2}(\lambda_2+\lambda_3)-\frac{1}{2}(\mu_1+\mu_2+\mu_4)
\end{equation}
\item[V]
    $\rho_1=\rho_2=\tau_2=0$: we choose as parameters
\begin{equation}	  \label{4.22}
\rho_i = \rho\xi_i, i \in \{1,2\}, \quad\tau_2 = \rho\xi_3
\end{equation}
and let $\rho \rightarrow 0$. The poles of $\sigma_1$ appear at
\begin{equation}	  \label{4.23}
\sigma_1 = \frac{1}{2}(\lambda_1+\lambda_2 -\mu_1-\mu_4+\mu_3) +s_1+s_2+n_0
\end{equation}
Provided the poles of $s_1, s_2$ are in $\bf{N}_0$, we find
\begin{equation}	  \label{4.24}
\gamma_5 = \frac{1}{2}(\lambda_1+\lambda_2-\mu_1-\mu_4+\mu_3)
\end{equation}
However, if we perform some of the integrations after	insertion of the delta function $\delta^{(0)}(\rho)$ corresponding	to the pole (\ref{4.29}) by (\ref{3.32}), we obtain a beta function with denominator $\Gamma(-2n_0)$. So these poles (V) cancel completely.
\item[VI]
     $\rho_1 = \rho_3 = \tau_2 = 0$:  We proceed as in the case (V) and get as pole positions
\begin{equation}   \label{4.25}
 \sigma_1=\lambda_1+\lambda_3+\frac{1}{2}(\mu_2+\mu_3-\mu_1-\mu_4)-\frac{1}{2}d+2(s_1+s_3)+n_0
\end{equation}
which, if the poles of $s_1, s_3$ are in $\bf{N}_0$, gives
\begin{equation}  \label{4.26}
\gamma_6 = \lambda_1+\lambda_3+\frac{1}{2}(\mu_2+\mu_3-\mu_1-\mu_4)-\frac{1}{2}d
\end{equation}
\item[VII]
      Finally, there is one $\Bbb{K}_2$ face: $\tau_2=\rho_1=\rho_2=\rho_3=0$: coordinates are
\begin{equation}  \label{4.27}
\rho_i=\rho\xi_i, \: i \in \{1,2,3\}, \quad \tau_2=\rho\chi_4, \quad \sum_i \xi_i = 1
\end{equation}
and we let $\rho \rightarrow 0$. We get the pole positions
\begin{equation}  \label{4.28}
\sigma_1=\frac{1}{2}(\lambda_1+\lambda_3-\mu_1-\mu_4)+s_1+s_3+n_0
\end{equation}
If $s_1, s_3$ have poles in $\bf{N}_0$, we obtain
\begin{equation}  \label{4.29}
\gamma_7=\frac{1}{2}(\lambda_1+\lambda_3-\mu_1-\mu_4).
\end{equation}
\end{enumerate}

The symmetry group of the graph Fig.7 is the same as that of Fig.4: $Z_2 \times Z_2$. It acts on the ${\lambda_i}$ as
\begin{alignat}{3} 
S_1(\lambda_i) &= \lambda_i, \: i \in {1,3},\quad S_1(\lambda_2) &= \lambda_4, \quad S_1(\lambda_4)&= \lambda_2 \label{4.30} \\
S_2(\lambda_i) &= \lambda_i, \: i \in {2,4},\quad S_2(\lambda_1) &= \lambda_3, \quad S_2(\lambda_3)&= \lambda_1 \label{4.31} 
\end{alignat} 
This rules out all $\gamma$'s, except $\gamma_1, \gamma_7$ and of course $\gamma_0=0$, which originates from the $\sigma_1$ poles of $\Gamma(-\sigma_1)$. Thus the AdS box graph has the same critical exponents as the CFT box graph Fig.6.

The poles in $s_4$ are all from $\Gamma(-s_4)$. The other variables $(s_1, s_2, s_3)$produce poles of the function $\Phi$ (4.4) that can be ordered in (triple) sequences
\begin{equation} \label{4.32}
\{(\nu_1+n_1, \nu_2+n_2, \nu_3+n_3), \, \nu_i \:\text{fixed}, \, n_i \in \bf{N}_0 \:\text{running}\}
\end{equation}
In the two tables below we list all possible triples $(\nu_1, \nu_2, \nu_3)$ and their connection ("origin") with the $\sigma_1$ singularities (I) - (VII).
The entries in the tables originating from the cases VI or VII are marked by (*). The corresponding $n_i$ runs over $\frac{1}{2}\bf{N}_0$ (not $\bf{N}_0$).
\begin{table}[htb]
\begin{tabular}{|c|c|c|c|}
	\hline
	origin&$\nu_1$&$\nu_2$&$\nu_3$\\
	\hline
        I&$0$&$0$&$0$\\
        II&$\frac{1}{2}(\mu_2-\lambda_1-\lambda_2)$&$0$&$0$\\
        II&$0$&$\frac{1}{2}(\mu_2-\lambda_1-\lambda_2)$&$0$\\
	II,IV,VII&$0$&$\frac{1}{2}(\mu_2-\lambda_1-\lambda_2)*$&$\frac{1}{2}(\mu_2+\mu_3-\lambda_1-\lambda_3)$\\
	II,IV,VII&$\frac{1}{2}(\mu_2-\lambda_1-\lambda_2)*$&$0$&$\frac{1}{2}(\mu_3+\lambda_2-\lambda_3)*$\\ 
	II,IV,VII&$\frac{1}{2}(\mu_2+\mu_3-\lambda_1-\lambda_3)$&$\frac{1}{2}(\lambda_3-\lambda_2-\mu_3)*$&$0$\\
	VI&$-\frac{1}{2}(\lambda_1+\lambda_3)+\frac{1}{4}d*$&$0$&$0$\\ 
	VI&$0$&$0$&$-\frac{1}{2}(\lambda_1+\lambda_3)+\frac{1}{4}d*$\\ 
	VII&$\frac{1}{2}(\mu_2+\mu_3-\lambda_1-\lambda_3)$&$0$&$0$\\
	VII&$0$&$0$&$\frac{1}{2}(\mu_2+\mu_3-\lambda_1-\lambda_3)$\\
	VII&$0$&$\frac{1}{2}(3\mu_2-d-\lambda_1-\lambda_2)$&$\frac{1}{2}(\mu_2+\mu_3-\lambda_1-\lambda_3)$\\
	VII&$\frac{1}{2}(3\mu_2-d-\lambda_1-\lambda_2)$&$0$&$\frac{1}{2}(\lambda_2-\lambda_3-2\mu_2+\mu_3+d)$\\
	VII&$\frac{1}{2}(\mu_2+\mu_3-\lambda_1-\lambda_3)$&$\frac{1}{2}(-\lambda_2+\lambda_3+2\mu_2-\mu_3-d)$&$0$\\
	\hline
\end{tabular} 
\caption{Sequences of $\left\{s_1, s_2, s_3 \right\}$ poles contributing to $u^{\gamma_1} F_1(u,v)$, $\gamma_1=\frac{1}{2}(\mu_2+\mu_3-\mu_1-\mu_4)$}
\end{table}
\begin{table}[htb]
\begin{tabular}{|c|c|c|c|}
	\hline
	origin&$\nu_1$&$\nu_2$&$\nu_3$\\
	\hline
	IV&$0$&$\frac{1}{2}(\mu_2-\lambda_1-\lambda_2)$&$0$\\
	VII&$0$&$0$&$0$\\
	VII&$0$&$\frac{1}{4}(-\lambda_1-2\lambda_2+\lambda_3+\mu_2-\mu_3)*$&$0$\\
	VII&$0$&$\frac{1}{2}(\lambda_1-\lambda_2+2\lambda_3+\mu_2-2\mu_3-d)*$&$0$\\
	\hline
\end{tabular} 
\caption{Sequences of $\left\{s_1, s_2, s_3 \right\}$ poles contributing to $u^{\gamma_7} F_7(u,v)$, $\gamma_7=\frac{1}{2}(\lambda_1+\lambda_3-\mu_1-\mu_4)$}
\end{table}
\section{Concluding Remarks}
\setcounter{equation}{0}
We have proved that for the box graphs of CFT$_d$ and AdS$_{d+1}$ supergravity, we obtain the same critical exponents, namely those which are determined from the "Cutkosky rule" with external lines included. We suggest that this behavior is also shown by other one-particle-irreducible graphs. Each critical exponent $\gamma_k$ belongs to one or more sequence of poles in the Mellin-Barnes parameters $(s_1, s_2, s_3)$, each of which is generated by a triple $(\nu_1, \nu_2, \nu_3)$ (see (\ref{4.32})), and each sequence contributes to the coefficient $c_{mn}^{(k)}$ in (\ref{2.6}). The larger the number of $\gamma_k, \nu_1, \nu_2, \nu_3$ that are nonzero, the smaller the number of remaining integrations. More details on this can be found in \cite{lhoffmann}.

\section*{Acknowledgements}
The authors thank A. C. Petkou for interesting discussions during the initial stages and one of us (W. R.) thanks the staff of the Werner-Heisenberg-Institut in Munich for their hospitality during the final stage of this work.


\begin{thebibliography}{99}

\bibitem{maldacena}
J. Maldacena, ``The large N limit of superconformal field theories and supergravity'', Adv. Theor. Math. Phys. {\bf 2} (1998) 
231-252, hep-th/9711200.
\bibitem{gubser}
S.S. Gubser, I.R. Klebanov and A. M. Polyakov, ``Gauge theory correlators from noncritical string theory'', Phys. Lett. {B428} (1998) 105, hep-th/9802109. 
\bibitem{witten}
E. Witten, ``Anti-de  Sitter space and holography'', Adv. Theor. Math.Phys. {\bf 2} (1998) 253-291, hep-th/9805028.
\bibitem{petersen}
J. L. Petersen, ``Introduction to Maldacena Conjecture on AdS/CFT, hep-th/9902131;
O.~Aharony, S.S.~Gubser, J.~Maldacena, H.~Ooguri and Y.~Oz, ``Large N field theories, string theory and gravity'', hep-th/9905111.
\bibitem{bilal}
A. Bilal, C.-S. Chu, ``A note on the Chiral Anomaly in the AdS/CFT correspondence and $\frac{1}{N^2}$ correction'', hep-th/9907106; ``Testing the AdS/CFT correspondence beyond the large $N$, hep-th/0003129.
\bibitem{petkou}
A. Petkou, K. Skenderis, ``A non-renormalization theorem for conformal anomalies'', hep-th/9906030.
\bibitem{bianchi}
M. Bianchi, S. Kovacs, ``Non-renormalization of extremal correlators in ${\cal{N}}=4$ SYM theory'', hep-th/9910016.
\bibitem{eden}
B. Eden, P.S. Howe, C. Schubert, E. Sokatchev, P. C. West, ``Extremal correlators in four-dimensional SCFT'', hep-th/9910150.
\bibitem{dhoker}
E. d'Hoker, S. D. Mathur, A. Matusis, L. Rastelli, ``The operator product expansion of ${\cal{N}}=4$ SYM and the $4$-point functions of supergravity'', hep-th/9911222.
\bibitem{arutyunov}
G. Arutyunov, S. Frolov, A. C. Petkou, ``The Operator Product Expansion of the Lowest Weight CPOs in ${\cal{N}}=4$ SYM$_4$ at Strong Coupling'', hep-th/0005182.
\bibitem{penati}
S. Penati, A.Santambrogio , D. Zanon, ``Two-point functions of chiral operators in ${\cal{N}}=4$ SYM at order $g^4$'', hep-th/9910197; ``Correlation functions of chiral primary operators in perturbative  ${\cal{N}}=4$ SYM'', hep-th/0003026.
\bibitem{liu}
H. Liu, ``Scattering in Anti-de-Sitter space and Operator Product Expansion'', hep-th/9811152; D. Freedman, S. D. Mathur, A. Matusis, L. Rastelli, ``Correlation functions in the CFT$_d$/AdS$_{d+1}$ correspondence'', hep-th/9804058.
\bibitem{hoffmann}
L. Hoffmann, A. Petkou, W. R\"uhl, ``Aspects of the Conformal Operator Product Expansion in AdS/CFT Correspondence'', hep-th/0002154.
\bibitem{dobrev}
V. K. Dobrev, G. Mack, V. B. Petkova, S. G. Petrova, I. Todorov, {\it Harmonic analysis on the $n$-dimensional conformal group and its applications to conformal quantum field theory}, Lecture Notes in Physics, Vol.63, Springer-Verlag, Berlin (1977).
\bibitem{cutkosky}
R. E. Cutkosky, J. Math.Phys.{\bf1} (1960), 429; see also C. Itzykson, J.-B. Zuber, {\it Quantum Field Theory}, Mc Graw-Hill, New York (1980): Section 6-3-4.
\bibitem{lang}
K. Lang, W. R\"uhl, ``The critical O($N$) $\sigma$-model at dimension $2<d<4$ and order $\frac{1}{N^2}$: Operator product expansions and renormalization'', Nucl.Phys. {\bf{B377}} (1992), 371.
\bibitem{symanzik}
K. Symanzik, ``On Calculations in Conformal Invariant Field Theories'', Lett. Nuovo Cim. {\bf 3} (1972), 734.
\bibitem{lhoffmann}
L. Hoffmann, University of Kaiserslautern, PhD Thesis, to appear.
\end{thebibliography}
\end{document}